\documentstyle[12pt]{article}

\newcommand{\bcn}{\begin{center}}
\newcommand{\beq}{\begin{equation}}
\newcommand{\beqn}{\begin{eqnarray}}
\newcommand{\ecn}{\end{center}}
\newcommand{\eeq}{\end{equation}}
\newcommand{\eeqn}{\end{eqnarray}}

 \def\lsim{\mathrel{\rlap{\lower4pt\hbox{\hskip1pt$\sim$}}
    \raise1pt\hbox{$<$}}}
\def\slash#1{\setbox0=\hbox{$#1$}#1\hskip-\wd0\hbox to\wd0{\hss\sl/\/\hss}}

\begin{document}

\rightline{KU-HEP-92-25}

\bcn
{\large\bf Systematic Analysis Method for Color
Transparency Experiments}\vspace*{2cm}

{\bf Pankaj Jain and John P. Ralston} \\

{\it Department of Physics and Astronomy \\
The University of Kansas \\
Lawrence, KS-66045-2151\\ }\vspace*{3cm}

{\bf ABSTRACT}
\ecn
\noindent We introduce a data analysis procedure for color transparency
experiments which is  considerably less model dependent than the transparency
ratio method. The new method is based  on fitting the shape of the A dependence
of the nuclear cross section at fixed momentum transfer to  determine the
effective attenuation cross section for hadrons propagating through the
nucleus. The  procedure does not require assumptions about the hard scattering
rate inside the nuclear medium.   Instead, the hard scattering rate is deduced
directly from the data. The only theoretical input  necessary is in modelling
the attenuation due to the nuclear medium, for which we use a simple
exponential law. We apply this procedure to the Brookhaven experiment of
Carroll et al and find  that it clearly shows color transparency: the effective
attenuation cross section in events with  momentum transfer $Q^2$ is
approximately
$40\ mb\ (2.2\ GeV^2/Q^2)$.  The fit to the data also supports  the idea that
the hard
scattering inside the nuclear medium is closer to perturbative QCD predictions
than is the scattering of isolated protons in free space. We also discuss
the application of our approach to electroproduction experiments.

\vfill\eject
\hspace*{3pc}

\noindent {\bf 1.} Color transparency [1, 2] is a theoretical prediction that
under certain
circumstances the strong  interactions may appear to be effectively reduced in
magnitude.  Brodsky and Mueller [1]  suggested measuring attenuation in a
nuclear target as a signal of color transparency. However, to  make a
quantitative measurement of the attenuation, one must have a value for the hard
scattering  rate.  In the absence of a normalization of the hard sub-process,
it is clear that only a combination of  the hard scattering rate and
attenuation rate is measured in an experiment.

\medskip

Theory at present cannot supply absolute normalizations for exclusive
processes, so the scattering  rate in an isolated hadron has been used
previously as a benchmark for comparison with the nuclear  target. The
"transparency ratio" T($Q^2$,A) was introduced by Carroll et al [3]:  it is the
ratio of a  cross section measured in the nuclear target to the analogous cross
section for isolated hadrons in  free space.  The transparency ratio is
convenient in experimental analysis because some of the  uncertainty in the
flux and other experimental uncertainties cancel.

\medskip

However, it has become clear that color transparency experiments cannot be
treated on such simple  terms.  The cross section for isolated pp$\to$pp
scattering shows oscillations which are model- independent evidence for
interference among competing amplitudes.  In dividing by the isolated
(free-space) pp cross section, one creates a certain bias, and indeed possibly
a
misleading energy  dependence in the transparency ratio [4].  On the other
hand, if one does not divide by the free  space cross section, but simply looks
at the $Q^2$ and A dependence of the nuclear target, what is the  magnitude of
the process to be compared to?  The experimental data in the nuclear target
actually  tracks an s$^{-10}$ naive power law behavior very well [5], but this
data represents a combination of the  underlying hard scattering and the
attenuation.  It is a model, and therefore not safe, to assume that  the hard
process goes like s$^{-10}$, where s is the c.m. energy -squared.  In fact, the
free space data  averages about s$^{-9.7}$, while apparently containing two
components,
and the whole question of  logarithmic effects presents a potentially large
correction not totally under theoretical control.

\medskip

The problem with interpreting the nuclear target data is that an overestimate
of the hard scattering  rate inside the nucleus might be partially compensated
by an underestimate of the nuclear  transparency.  If one chooses, for example,
to ``normalize" the nuclear hard scattering rate to Z times  some average of
the
pp$\to$pp data, one would be assuming that such hard processes occurred
equally
often in free space as in the nuclear medium.  However, the filtering away of
the oscillations  indicate this is not true.  In fact, the assumption probably
overestimates the true rate of ``mini- events" occurring inside the nucleus.
In
general,  when one studies the ratio of nuclear to free space  quantities even
at any fixed energy, {\em one is likely to be comparing two unlike processes},
and opening  the door to a possibly unscientific procedure and confusion.

\medskip

This problem has gotten the most attention in the hadron-hadron case.  However,
the essential  arbitrariness of choosing to divide a nuclear quantity by a
free-space, isolated hadron quantity can  create uncertainty in all color
transparency experiments.   Electroproduction is not immune to the  problem,
because the free space form factors are not totally established as short
distance objects,  and their interpretation is certainly model dependent.
Suppose, for example, that for a certain  kinematic point 50\% of the
scattering
rate in free space were due to processes which do not occur in  a nuclear
target.  When one interpreted the transparency ratio one might then
overestimate the  scattering rate in the nuclear target by a factor of two. By
fitting the data with this assumption, the  survival rate due to attenuation
would be underestimated by 50\%.  This simple difficulty is quite a  serious
issue in electroproduction, because a hadron crosses the target only once.
Then, as we show  below, a measurement of an attenuation cross section with
limited statistics can be completely upset  by a change in normalization.
However, with good enough statistics one can actually separate the  hard
scattering rate and attenuation rate in the new way we describe.

\bigskip

\noindent {\bf 2.} In this paper we outline a data analysis procedure which is
much less model
dependent than the  transparency ratio method.  We never assume the
transparency ratio measures attenuation, because  the transparency ratio is a
ratio between two unknowns.  We recommend a global fitting procedure  to use
all the information in sets of data, with a free parameter setting the
normalization of the  experiment.   A crucial point is that the shape of the
cross section with A for a set of nuclear targets  contains a great deal of
information, and is the best way to deduce the effective attenuation cross
section.  The normalization of the data is then determined empirically, without
excessive model- dependence.  The normalization does not have to be set using
isolated hadron events, and in fact the  empirically determined normalization
generally will disagree with the normalization assumed from  comparing to such
``free space" data.  This is progress:  we can use the data itself to learn
about the  differences between the nuclear and free space cases.

\medskip

We first consider the BNL experiment and its A dependence at fixed beam energy.
The  experimental data as plotted by Heppelmann [5] is shown in Fig. (1).  We
analyze the cross section  in the nuclear target, dispensing with the
transparency ratio.   Recall that the events from the nuclear  target are taken
at the kinematic point  corresponding to $90^o$ cm scattering;
the beam energy $E$
then  determines $s\ =\ 2mE\ +\ 2m^2$.  We let $Q^2$ be the momentum transfer,
$Q^2= -t = s/2-2m^2$.  So long as  the amplitudes have a decent factorization,
we can assume that the cross section is a product of a  hard scattering
probability and an
attenuation probability.  We find the attenuation from interactions  with cross
section $\sigma_{eff}$ for a proton leaving a nucleus via a straight line path.
 We
assume the hard  scattering occurs at random over the volume of the nucleus.
We then fit the data to a Monte-Carlo event generator\footnote{We thank
Steve Hepplemann for lending us the Monte Carlo code used in Ref. (5).} in
which
$\sigma_{eff}$ is a free
parameter, using the parametrizations of the nuclear density  given in [6].
The Monte Carlo propagates the incoming proton to the hard scattering point,
and follows the outgoing protons along all kinematically allowed random
paths. At each point of each path the loss rate is proportional to the
flux, the number density and $\sigma_{eff}$.
For reference we have plotted the nuclear number densities in Fig.(2);  there
is  considerable variation at the edges.  It is doubtful that including more
detailed structure of the  nucleus such as correlations would be justified
within the experimental and theoretical errors.  Typical results of the
Monte-Carlo are shown in Fig.(3), where the dashed lines are smooth fits to
the Monte-Carlo output.

\medskip

Since the Monte-Carlo is a ``black-box" a useful consistency check is an
analytic representation.   The simplest thing is to calculate the survival
$P_p$ of one proton crossing on a straight line, assuming uniform nuclear
density $\rho\ =\ (1/6)\ fm^{-3}$ and nuclear radius $(1.2\ fm)\ A^{1/3}$.
An approximate survival probability is then given by cubing the result
and renormalizing.  The volume-averaged survival probability
$P_{ppp}(\sigma_{eff},\ A)$,
 which we  call the ``survival factor", is then given by
\beqn
P_{ppp}(\sigma_{eff},\ A)& \cong &1.9\biggl({3\over
4n\sigma_{eff}1.2A^{1/3}}+{3\over 8n^3\sigma_{eff}^31.2^3A}\nonumber \\
&\times&\bigg\lbrack\exp(-2n\sigma_{eff}1.2A^{1/3})-1+2n\sigma_{eff}1.2A^{1/3}
\exp(-2n\sigma_{eff}1.2A^{1/3})\bigg\rbrack\biggr)^3\nonumber \\
\eeqn
where $\sigma_{eff}$ is the attenuation cross section in units of $fm^2$. With
 the renormalization
 factor 1.9 this  primitive approximation works surprisingly well for
$A\geq7$ and the analysis of the data agrees within  errors.  We do not think
that
few nucleon effects are well accounted for by either method, so we  restrict
ourselves to $A\geq7$. For the case of an electroproduction experiment, there
is
only one proton  which must cross the nucleus, and we use a survival factor
$P_p$
 which is ($P_{ppp}/1.9)^{1/3}$.   A main  objective of color transparency
experiments
is to determine the survival factor $P(\sigma_{eff},\ A)$, or  equivalently the
attenuation cross section $\sigma_{eff}$ as a function of energy or momentum
transfer.

\bigskip

\noindent {\bf 3.} In step (1) of our procedure,  for each energy we fit the
shape of the
nuclear cross section to  determine a free normalization $N(Q^2)$ and an
attenuation cross section $\sigma_{eff}(Q^2)$ .  For the BNL  experiment this
is:
\beq
s^{10}{d\sigma_{_A}/dt\over Z}=N(Q^2)\; P_{ppp}(\sigma_{eff}(Q^2),\; A)
\eeq
The factors of $s^{10}$ and 1/Z are simply definitions to take out some typical
orders of magnitude -  their usage will introduce no bias.  Since
$\sigma_{eff}(Q^2)$
and
the normalization N($Q^2$) are free, the fit to  the data may or may not show
that
these parameters vary with energy.  Of course, finding a  variation with energy
of  $\sigma_{eff}$ was the goal of color transparency, but at this stage the
fitting
is an  objective, empirical procedure.  In the BNL data, for example, the best
$\chi^2$ fit to the shape of the  nuclear cross section is obtained by

\beqn
\sigma_{eff}(E=6\ GeV)=17^{+5}_{-3}\ mb;\ \ \  N(E=6\ GeV)&=&(5.4\pm0.4)\zeta\;
\hskip .4in \chi^2=0.28 \nonumber \\
\sigma_{eff}(E=10\ GeV)=12^{+5}_{-4}\ mb;\ \ \ N(E=10\
GeV)&=&(3.3\pm0.4)\zeta\;
\hskip .4in \chi^2=0.53\nonumber
\eeqn
where $\chi^2 = \Sigma[(y_i - d_i)/\Delta di]^2$, $d_i$ are the data points,
$\Delta d_i$ is the error in $d_i$  and $y_i$ are the theoretical values
calculated using
the Monte-Carlo, and $\zeta = 5.2\times10^7$ mb GeV$^{18}$ is a constant
containing the  overall normalization of the  cross section.  Note that
$\chi^2$ has not yet been divided by the number of  degrees  of freedom:  this
is discussed below.  The fits for 6 GeV and 10 GeV are shown in Fig.  (4a) and
Fig. (4b) respectively; the low $\chi^2$ values indicate rather good fits to
five data points. (The  fit does not change significantly if the data point of
Lithium is deleted: $\sigma_{eff}$(E = 6 GeV) = (19$^{+17}_{-6}$)mb and
$\sigma_{eff}$(E
= 10 GeV) = (13$^{+7}_{-4}$) mb in that case.)  At these beam energies we
assign
the  nominal values of $Q^2$ = 4.8 GeV$^2$ and 8.5 GeV$^2$, respectively - this
is  discussed below.  The  important thing to notice is the curvature of the
plots, which is the main determiner of the value of  $\sigma_{eff}$ -  its
effect is
easily separated from the normalization constant\footnote{If the data is
divided by some arbitrary value, then the shape of the A  dependence cannot be
fit as well,  and  $\sigma_{eff}$  is forced to whatever value is needed to
reproduce the normalization of the data}. To illustrate this point we have
also included in Figs.(4a, 4b) the best fits with various permutations of
$\sigma_{eff}$. When $\sigma_{eff}$ is  constrained to the wrong values of 12
mb and 36
mb at $Q^2$ = 4.8 GeV$^2$, the $\chi^2$ values for the best  arbitrary
normalization rise by an order of magnitude to 3.1 and 5.3 respectively.
Similarly, when  $\sigma_{eff}$ is constrained to the wrong values of 17 mb and
36
mb at $Q^2$ = 8.5 GeV$^2$, the $\chi^2$ values for the  best arbitrary
normalization rise to 1.6 and 5.3 respectively.  Although fixing $\sigma_{eff}$
reduces the  number of parameters, the $\chi^2$ per degree of freedom rises to
(0.78, 1.3, 0.4, 1.3) for the four cases  above, from four to thirteen times
bigger than the value of 0.093 of our best fits.  This means that  the best fit
is objectively better that the others by a significant statistical measure;
the proposal to fit  the data with a 36 mb cross section is highly disfavored
compared to the best fit.

\medskip

Having fit the data's A dependence, the fit can be examined at fixed A to find
the survival factor as a  function of $Q^2$ for fixed A.  The BNL data was
taken
only at two energies for a good range of A, so  we have $N(Q^2)$ and
$\sigma_{eff}(Q^2$) at
two points.  However, each fixed beam energy actually probes a  range of  true
event cm energies and $Q^2$ because of Fermi motion.  What can be deduced from
an
event at a single beam energy depends strongly on how many momenta are measured
among the  outgoing protons.  With enough momentum resolution and variables
measured, the Fermi motion is  actually measured independently, and this was
done well enough at BNL.  This procedure produced  data at intermediate
energies for the Aluminum target.  For the other targets, the data was
re-binned  to increased its statistical significance.  This causes  a potential
``binning problem"  because the  spread due to Fermi motion effect is not
exactly symmetric, but inspection of the Aluminum data  shows that this should
not be a big effect.  We therefore accept the data as it is given at the
re-binned points.

\medskip

We can make a function for $N(Q^2)$ which interpolates with $Q^2$
smoothly between
the two endpoints  where $Q^2$Jwas reported.  Then inverting (2) we have
\beq{(Q^2/4.8 GeV^2)^{0.86}\over 5.4\zeta}{s^{10}\over
Z}{d\sigma_A\over dt}=P_{ppp}(\sigma_{eff}(Q^2),A)\eeq
This is an implicit relation for $\sigma_{eff}$($Q^2$). We also have the
Aluminum data
at intermediate values of  $Q^2$ which can be used to determine
$\sigma_{eff}$($Q^2$).  The
resulting best consistent fit for $\sigma_{eff}$($Q^2$) is shown  in Fig. (5).
We
emphasize
that future data would be most useful if reported for all A and all $Q^2$ in
which case determining $\sigma_{eff}$($Q^2$) is more direct.  With
$\sigma_{eff}$($Q^2$)
determined the
survival probability  itself as a function of $Q^2$ is plotted in Fig. (6). The
survival factor would be flat with $Q^2$ if one used  a traditional Glauber
model.
The fact that the survival probability rises with $Q^2$ is clear evidence for
discovery of color transparency.  It has gone unnoticed in studies involving
the transparency ratio,  because the assumed normalization in that case skews
the analysis.

\medskip

The transparency ratio, of course, has a certain usefulness, because any
experimental uncertainty in  the overall normalization drops out.  In fact,
the overall normalization drops out of our method,  also:  one does not need to
know the normalization to determine $\sigma_{eff}$.  If there is one overall
uncertainty
in the normalization, it also simply scales out,  and does not affect the
procedure.  Thus  we are not asking for absolute cross sections, which are
usually a difficult thing.  Our method could  be applied directly to the
transparency ratio if desired - the fitting of the shape of the A-dependence
is exactly the same.  However, in that case the normalization factor $N(Q^2)$
will
reflect the $Q^2$  dependence of whatever normalization is chosen.

\medskip

The effective attenuation cross section $\sigma_{eff}$ should be a universal
quantity
which can be compared  from experiment to experiment.  The theoretical basis
for this is factorization between the hard  scattering and the nuclear
propagation factors.  The hard scattering is an abstract object- an integral
kernel - which is independent of the target, but when convoluted with different
wave functions in   different targets produces different effects.  An
approximate fit to the attenuation cross section from the fit is
\beq
\sigma_{eff}(Q^2)= 40 mb (2.2 GeV^2/ Q^2)\pm 5 mb
\eeq
for 4.8 $GeV^2< Q^2 <
8.5 GeV^2$.  The decrease with $Q^2$ of the attenuation cross section coincides
with the rate predicted theoretically on the basis that the cross section goes
like the transverse  separations $b_T^2$ of the participating quarks, and
that the
region of important $b_T^2 = 1/Q^2$[4, 7].   However, the scale in the
numerator of
2.2 GeV$^2$ was not predicted.  To put the scale in context,  it  says that the
perturbative QCD ideas are beginning to apply for Q$\geq$1.5 GeV.  This is
another
way  to present the conclusion that color transparency was actually observed in
the BNL experiment.

\medskip

A final consistency check involves looking at the the normalization factor.
After taking out the  nuclear attenuation effects, according to the
perturbative treatment the $Q^2$ dependence of $N(Q^2)$ is  due to the hard
scattering process.  We have found that $N(Q^2)$ decreases relative
to $s^{-10}$,
meaning  that the hard scattering rate in the nuclear target is decreasing
faster than the naive quark counting  model prediction.  In perturbative QCD,
however, the quark-counting  prediction is modified,  due  to the running
coupling $\alpha_s(Q^2/\Lambda^2_{qcd})$ and scaling of distribution
amplitudes.  The perturbative
QCD prediction goes like $\alpha_s^{10}$ because there are five gluons in the
amplitude:
\beq d\sigma/dt_{pQCD} = (\alpha_s(Q^2/\Lambda_{qcd}^2))^{10}s^{-10}f(t/s)\eeq
It becomes interesting to compare the form including powers of $\alpha_s$ with
the hard
scattering rate in  the nuclear target. (The $Q^2$ dependence of
$\alpha_s^{10}$ causes
serious disagreement with the data for  isolated pp scattering in free space
with $\Lambda_{qcd} \approx$ 100 MeV.)  The reason for consulting the
perturbative prediction
is to see whether the nuclear target has filtered the events down to something
like the shortest distance component.  Although (5) appears naive it is
perfectly adequate because of  the usual ambiguity in the choice of scale.  We
must generate a range of reasonable theoretical  predictions by choosing the
$Q^2$
of $\alpha_s$ in a typical range (-t/2) $< Q^2 <$ (-1.5 t).  To improve on this
theoretically requires a calculation of next to leading logarithms.  The
effects of certain leading  corrections due to anomalous dimensions fall within
the range;  for example, the asymptotically  large $Q^2$ prediction amounts to
multiplying (5) by  $[\log(Q^2/\Lambda_{qcd}^2)]^{-8\gamma_1}$, where
$\gamma_1$
is the smallest  anomalous dimension [8].

\medskip

For comparison with the nuclear target we form the ratio of the global
$s^{-10}N(Q^2)$ to the pQCD  predictions (5) and plot the result
as solid lines in
Fig. (7). (The asymptotic prediction is also  illustrated as a dashed line to
show that it falls within the region of scale ambiguity of the pQCD
predictions.  For comparison we show s$^{-9.7}$, which is well outside the
range of
the perturbative  predictions.)  The solid lines of
$s^{-10}N(Q^2)/(d\sigma/dt_{pQCD}(Q^2)$)
are quite flat, showing good  agreement of the BNL nuclear hard scattering data
with QCD.  For the Aluminum target we have  several data points at several
$Q^2$
which also fall fairly well within the band of perturbative  predictions.
These are rather spectacular results, but we emphasize that they should be
viewed with  caution because the BNL experiment was a pioneering one.  If it is
confirmed by upcoming higher  precision data,  they will lead a great deal of
strength to the idea that QCD is cleaner after filtering in  a nuclear target.

\bigskip

\noindent {\bf 4.} The procedure used here to analyze the Brookhaven data can
be applied
directly to quasi-elastic  electroproduction.  It is important to approach the
analysis steps in the right order.  Step (1) is to fit  the shape of the A
dependence of data with the attenuation cross section $\sigma_{eff}$($Q^2$) as
well
as the
normalization $N(Q^2)$ treated as unknown parameters for each $Q^2$.  For this
step
the data can be  either in the form of the cross section in the nuclear medium
or the color transparency ratio, since  the normalization is a separate
parameter.  Following this procedure for several values of $Q^2$ will
determine
the attenuation cross section $\sigma_{eff}$($Q^2$) as well as the
normalization $N(Q^2)$ as
functions of $Q^2$. The survival factor $P_p(\sigma_{eff}(Q^2$), A), which is
the essential
measure of color transparency, is then  obtained as a function of $Q^2$ and A
by
reinserting $\sigma_{eff}$ ($Q^2$).

\medskip

For consistency, quasi-exclusive electroproduction experiments such as the
SLAC-NE18   experiment should see the same attenuation cross sections at the
same $Q^2$.  With the fit to the BNL  data we calculate the survival factor
$P_p(\sigma_{eff}(Q^2$), A). The results are plotted in Fig. (8) as a function
of A and in
Fig. (9) as a function of $Q^2$.  Inspecting the log-scale Fig. (8), one sees
almost parallel  curves, although the values of $\sigma_{eff}$ in the plot vary
from 36
mb to 12 mb.  Thus one-particle quasi- exclusive electroproduction has a
relatively low discriminating power in determining the attenuation  cross
section. With limited statistics, an electroproduction experiment can certainly
determine a  curve in the hard scattering rate-attenuation cross section plane;
with good statistics, the curvature  of the A dependence can independently fix
$\sigma_{eff}$ and the hard scattering rate.

\medskip

{}From N($Q^2$) one can extract an effective form factor for the proton in a
nuclear
target. The form  factors in the nucleus are objects of great intrinsic
interest.  The comparison with the free space  form factors is also extremely
interesting.  Based on the idea that perhaps not all of the free space  form
factor is really due to short distance, our expectation is that the form
factors extracted from a  nuclear target  such as iron may be normalized by
50-70\% of the free space ones around $Q^2$ = 6  GeV$^2$.   The pQCD prediction
for dependence on momentum transfer is
\beq G_M(Q^2)_{pQCD} = {\rm const} (\alpha_s(Q^2/\Lambda_{qcd}^2))^2/ (Q^4)\eeq
Thus, if the pattern we observed in the BNL experiment is repeated, the form
factors in the nuclear  target should show a dependence on momentum transfer
that falls significantly more rapidly than  the naive quark-counting rule. The
effects of the anomalous dimension are again smaller than the  effects of scale
ambiguity and can be dropped, assuming one varies the $Q^2$ scale of $\alpha_s$
over a
typical  range $(-Q_o^2/2) < Q^2 < (-1.5 Q_o^2)$, where $Q_o^2$ is the photon
momentum
transfer.  This gives us a  band of theoretical predictions. In addition there
are higher twist effects, as evidenced by the dipole  fit to the free space
data $G_M(Q^2)_{data} =$ const./($Q^2+ .7$ GeV$^2)^2$.  Assuming somewhat
arbitrarily  that
this cancels out,  in Fig. (10) we plot  the product
\beq T(Q^2, A) =  {\rm const}. [\alpha_s(Q^2/\Lambda_{qcd}^2)/ \alpha_s(1
GeV^2/\Lambda_{qcd}^2)]^4 P_p(\sigma_{eff}(Q^2), A)\eeq
as a function of $Q^2$ for various A, and with arbitrary ``const".  The ratio
calculated in (7) is a very  rough\footnote{In general the
transparency ratio depends on the specific cross sections used for comparison
and there is no  universal agreement or convention established.} ``transparency
ratio" using the guess that
the hard scattering rate in a large enough nucleus  is perturbative enough to
show the running coupling. Then we find that the $Q^2$ dependence of (7) is
practically flat (Fig. (10)). Since we emphasize data analysis here we do not
attempt to estimate the  constant, which would represent the percentage of form
factor (squared) that would be filtered away  by the nuclear medium, and also
depends on the denominator of $\alpha_s$ in (7).  The constants cancel out  in
ratios
for A$>>$1; our calculation of the ratios varies from Fe/C = 0.69, Au/ C = 0.48
at $Q^2$ = 3  GeV$^2$ to  Fe/C = 0.78, Au/ C = 0.60 at $Q^2$= 6.8 GeV$^2$.  All
of these
are estimates based on rough  ideas.  The transparency ratio does not have the
same information as the separate extractions of the  hard scattering rate and
survival probability, and is not a substitute for the simpler and more
systematic method we are presenting.

\bigskip

\noindent {\bf 5.} In conclusion, we remark that the experimental determination
of color
transparency seems to be  more subtle than was originally supposed.  The basic
issue is that exclusive hard scattering contains  many unknowns, which has made
unraveling the data challenging.  However, the original concept of  color
transparency was to use a nucleus as a test medium to observe and compare
scattering with  the free space processes.  We believe this concept is sound
and is beginning to yield results.  Our  method uses more of the information
from the experimental data and we have been able to deduce a  great deal.  The
BNL data strongly indicates an attenuation cross section that decreases with
$Q^2$
at  the same rate as perturbatively predicted.  The same data strongly
indicates a normalization of the  hard scattering rate that decreases faster
than the free space rate, and is in rather good agreement  with perturbative
theory. These two results are not totally surprising.  Indeed, if we had much
faith  in the perturbative prediction's dependence on $Q^2$ due to many powers
of
$\alpha_s$, then the interplay of a  falling cross section and rising survival
might
have been obvious in the first look at the data.  The  goal is to have
sufficiently strong data and systematic data analysis so that no faith is
required.

\medskip

With the current data everything indicates that the new method is a powerful
one which can be  productively applied to color transparency experiments in the
future.  We eagerly await new data, in  the hopes that it will confirm our
conclusion that color transparency was actually observed by  Carroll et.al.

\bigskip

{\bf Acknowledgements:}    This work has been supported in part by the DOE
Grant No.
DE-FG02- 85-ER-40214.A008.  We thank Stan Brodsky, Alan Carroll, Steve
Heppelmann, Byron Jennings,  and Bob McKeown for useful comments.
\newpage

\noindent {\bf Figure Captions}

\bigskip

\baselineskip=12pt
\parindent=0pt
Fig.(1)  The A dependence of the data of Carroll et al. as reported by
Heppelmann (Ref. (5)).    There is a clear difference in curvature of the A
dependence between 6 GeV and 10 GeV indicating  a smaller attenuation cross
section at 10 GeV.  Due to oscillations in the denominator, the relative
normalization of the transparency ratio at the two energies is a separate
issue.

\medskip

Fig. (2)  Plot of nuclear number densities $\rho$(r) from Ref.(7) used in the
Monte
Carlo calculation.   The elements Li, C, Al, Cu, and Pb correspond to nuclear
A = 7, 12, 27, 64, and 207.

\medskip

Fig. (3) Smooth fits (dashed lines) to the Monte-Carlo calculation (solid
lines) of the survival  probability $P_{ppp}(\sigma_{eff}$, A) as a function of
$\sigma_{eff}$ for
various A.

\medskip

Fig. (4)  Fitting the A dependence of the nuclear cross section with
attenuation cross sections $\sigma_{eff}$ =  12, 17, and 36 mb at fixed
momentum
transfer. (a) $Q^2= 4.8$ GeV$^2$. (b) $Q^2= 8.5$ GeV$^2$.  The solid  line
shows the best
fit.  For comparison the dashed lines show constrained fits using permutations
of $\sigma_{eff}$ and best normalizations.

\medskip

Fig. (5)  The attenuation cross section $\sigma_{eff}$ extracted from the BNL
data.
Solid line: best fit  including the intermediate $Q^2$ data points reported for
Aluminum.  Dashed line:  a 36 mb strong  interaction cross section for
comparison.  The two data points are the global $\sigma_{eff}$ values extracted
from
Fig.(4); the uncertainty corresponds to the standard definition, producing a
change of the fit  by one unit of $\chi^2$.

\medskip

Fig. (6)  The survival probability $P_{ppp}$ extracted from the BNL data as a
function of $Q^2$ for various  A.  The increase with $Q^2$ is evidence for
observation of color transparency.

\medskip

Fig. (7)  Comparison of the fit to the hard scattering rate N($Q^2$) to pQCD
prediction, as given in  Eq. (5), including scale ambiguity.  The ratio
s$^{-10}$N($Q^2$)/(d$\sigma$/dt$_{pQCD}$) is plotted as solid lines;  a  flat
curve indicates
agreement.  The lower line uses the scale of the running coupling as ($Q^2$ =
0.5(-t)) in calculating d$\sigma$/dt$_{pQCD}$.  The upper line uses
$\alpha_s(Q^2$ = 1.5(-t)).
Data points are for the  complete data set (box symbol) and the Aluminum data
(no symbol). The long dashed curve  represents s$^{-9.7}/(d\sigma/dt_{pQCD}$)
and the short
dashed curve is the asymptotic prediction [8] for $d\sigma/dt_{pQCD}$ using
$\alpha_s(Q^2 = -t)$ and divided by the right hand side of Eq. 5.

\medskip

Fig. (8)  Prediction of the survival probability $P_p$ in quasi-exclusive
electroproduction versus A  using $\sigma_{eff}$($Q^2$) extracted from the BNL
data.  Note
that a change in normalization can nearly  compensate a change in
$\sigma_{eff}$.

\medskip

Fig. (9)  Prediction of the survival probability $P_p$ in quasi-exclusive
electroproduction versus $Q^2$  using $\sigma_{eff}$($Q^2$) extracted from the
BNL data.  A = 12, 56, 197 for elements C, Fe, Au.

\medskip

Fig. (10)  $Q^2$ dependence of a ``transparency ratio" consisting of
$\alpha^4_s(Q^2) P_p$ for
Fe (Eq. 7) under  the assumption that the nuclear form factor goes like
$\alpha_s^2$.
Lower and upper lines show the range of  theory predictions from varying the
choice of scale $Q^2$ of $\alpha_s$ from $(-0.5 Q_o^2) < Q^2 <(-1.5 Q_o^2)$,
where
$Q_o^2$ is the
photon momentum transfer.  Vertical scale depends on overall normalization of
hard scattering rate in nuclear medium which has not been specified.

\newpage
\noindent {\bf References}
\bigskip
\begin{enumerate}
\item S. J. Brodsky and A. H. Mueller, Phys. Lett. B {\bf 206}, 685 (1988), and
references therein.

\item S. J. Brodsky and G. F. de Teramond, Phys. Rev. Lett. {\bf 60}, 1924
(1988);
N. N. Nikolaev  and B. G. Zakharov, Z. Phys. {\bf C49}, 607 (1991); J. Botts,
Phys.
Rev. D {\bf 44}, 2768(1991); J.  P. Ralston and B. Pire, Nucl. Phys. A {\bf
532}, 155c
(1991) and Phys. Lett. {\bf B256,} 523 (1991);  H. Borel, S. Fleck, J.
Marroncle, F.
Staley, and C.Vallet, Nucl. Phys. A {\bf 532}, 291c (1991);  C. E. Carlson and
J.
Milana, Phys. Rev. D {\bf 44}, 1377 (1991);  J. P. Ralston, Phys. Lett. B  {\bf
269}, 439
(1991); P. Jain, J. Schechter and H. Weigel, Phys. Rev. D {\bf 45}, 1470
(1992); O.
Benhar, A. Fabrocini, S. Fantoni, V. R. Pandharipande, and I. Sick, Nucl. Phys,
A {\bf 532},  277c(1991); J. Hufner and M. Simbel, Phys. Lett. B {\bf 258}, 465
(1991); B.
Z. Kopeliovich  and B. G. Zakharov, Phys. Lett. B {\bf 264}, 434 (1991); Phys
Rev D
{\bf 44}, 3466 (1991); J. M.  Eisenberg and G. Kalbermann, preprint \#
PRINT-92-0126
TEL-AVIV); H. Heiselberg,  G. Baym, B. Blaettel, L. L. Frankfurt and M.
Strikman, Phys. Rev. Lett. {\bf 67}, 2946 (1991);  T.-S. H. Lee and G. A.
Miller,
Phys. Rev. C {\bf 45}, 1863 (1992); G. R. Farrar, H. Liu, L.  Frankfurt and M.
Strikman, Phys. Rev. Lett. {\bf 61}, 686 (1988); B. Jennings and G. Miller,
Phys.
Lett. {\bf B236}, 209 (1990); Phys. Rev. D {\bf 44}, 692 (1991); Phys. Lett.
{\bf B274}, 442
(1992);  S. Gardner, CEBAF preprint \# CEBAF-PR-92-002.

\item  A. S. Carroll et al., Phys. Rev. Lett. {\bf 61}, 1698 (1988).

\item J. P. Ralston and B. Pire, Phys. Rev. Lett. {\bf 61}, 1823 (1988).

\item S. Heppelmann, Nucl. Phys. B (Proc. Suppl.) {\bf 12}, 159 (1990).

\item H. De Vries, et al , Atomic Data and Nuclear Data Tables, {\bf 36}, 495
(1987).

\item S. Nussinov, Phys. Rev. Lett {\bf 34}, 1286 (1975); F. E. Low , Phys.
Rev.
{\bf D12}, 163, (1975);  J.  Gunion and D. Soper, Phys. Rev. {\bf D15}, 2617,
(1977).

\item S. J. Brodsky and G. P. Lepage, Phys. Rev. {\bf D22}, 2157 (1980).

\end{enumerate}
\end{document}